# Solvency assessment within the ORSA framework: issues and quantitative methodologies

Julien Vedani

Université de Lyon, Université Claude Bernard Lyon 1, Institut de Science Financière et d'Assurances, 50 Avenue Tony Garnier, F-69007 Lyon, France
julien.vedani@etu.univ-lyon1.fr

PhD Student – Milliman Paris
julien.vedani@milliman.com

Laurent Devineau

Université de Lyon, Université Claude Bernard Lyon 1, Institut de Science Financière et d'Assurances, 50 Avenue Tony Garnier, F-69007 Lyon, France
laurent.devineau@isfaserveur.univ-lyon1.fr

Head of R&D – Milliman Paris
laurent.devineau@milliman.com

## Abstract

The implementation of the Own Risk and Solvency Assessment is a critical issue raised by Pillar II of Solvency II framework. In particular the Overall Solvency Needs calculation left the Insurance companies to define an optimal entity-specific solvency constraint on a multi-year time horizon. In a life insurance society framework, the intuitive approaches to answer this problem can sometimes lead to new implementation issues linked to the highly stochastic nature of the methodologies used to project a company Net Asset Value over several years. One alternative approach can be the use of polynomial proxies to replicate the outcomes of this variable throughout the time horizon.

Polynomial functions are already considered as efficient replication methodologies for the Net Asset Value over 1 year. The Curve Fitting and Least Squares Monte-Carlo procedures are the best-known examples of such procedures. In this article we introduce a possibility of adaptation for these methodologies to be used on a multi-year time horizon, in order to assess the Overall Solvency Needs.

## 1. Introduction

At the heart of the Solvency II directive, Article 45 introduces the guidelines of the Own Risk and Solvency Assessment (ORSA) process. The ORSA framework leads companies towards a better understanding and an optimal management of their risk profiles, consistent with their strategic choices. Its implementation requires a thorough analysis of both short and long term risks. One major challenge of the introduced framework is the definition of the Overall Solvency Needs, which corresponds to the capital level required in order to satisfy a solvency constraint resulting from the firm's strategic choices and generally speaking from the firm's risk appetite. Its practical assessment should be made through a prospective analysis of the Net Asset Value ($NAV$) funding need. This implies the choice of a relevant time horizon, in order to take into account the overall strategic business plan of the undertaking, and the choice of an approach of the multi-year solvency. This latest notion is particularly difficult to apprehend. The solvency concept is indeed well-known in its regulatory sense, related to the first pillar of the directive and to a 1-year time horizon. But it must be generalized to a more economic framework in order to optimally fit the strategic issues of insurance companies.

The notion of multi-year ruin has already been studied theoretically through various ruin models, as presented in Assmussen and Albrecher (2010), or in Rullière and Loisel (2004). However, the various theoretical assumptions leading to semi-closed formulas do not fit the Overall Solvency Needs simulatory framework and the great complexity of general life insurance products. The first objective of this paper, developed through Section 2, is to propose a practical formalization of the multi-year solvency concept. We present three main approaches of this notion, depending on the underlying goals and on the considered risk variable ($NAV$ or Solvency Ratio). These approaches can be related to three kinds of metrics leading to different implementation constraints.

The use of the most complicated constraints presented in Section 2 introduces great implementation issues. Indeed, the prospective analysis of the $NAV$ and of the Solvency Ratio requires the user to be able to project values of these underlying variables through a multi-year time horizon. As presented in Section 3, the Nested Simulations may be a relevant methodology to achieve the necessary projections. However, the highly stochastic nature of the methodology leads to very complex and time-consuming implementations. Several operational studies have been proposed in order to accelerate the estimation of high quantile values of the $NAV$, for a single-period time horizon, see for example Devineau and Loisel (2009) or Planchet and Nteukam (2012). In the multi-year framework, a first approach based on closed formulas has been studied by Bonnin and al. (2012). To our knowledge, no acceleration methodology has been adapted to the Overall Solvency Needs assessment framework without needing relatively strong model assumptions. In order to enable the use of the proposed metrics, we develop implementation alternatives based on multi-year adaptations of the polynomial proxies procedures already known as Curve Fitting or Least Squares Monte-Carlo, as presented respectively in Algorithmics (2011) and Barrie & Hibbert (2011).

In Section 4, we formalize the mathematical basics of these proxy methodologies. This enables us to prove the convergence of both procedures and to propose a formula in order to compare the asymptotic efficiency of both Curve Fitting and Least Squares Monte-Carlo through a multi-year horizon. In the last section, we present the results obtained after having implemented the methodologies proposed in Section 3, on a standardized life insurance product.

## 2. Single-period and multi-year solvency

### 2.1. Single-period solvency

The notion of 1-year horizon solvency, in its regulatory sense, is defined by the directive's first pillar. It is based on the $VaR_{99.5\%}$ risk measure. Being solvent means having enough own funds to be able to avoid economic bankruptcy over 1 year with a 99.5% threshold.

Let $NAV_t$ be the $NAV$ at time $t \geq 0$, $SCR_0$ be the initial 1-year regulatory capital, the Solvency Capital Requirement ($SCR$), and $\delta_1$ be the discount factor at the end of the first period.

The regulatory solvency constraint is denoted

$$\mathbb{P}(NAV_1 \geq 0) \geq 99.5\%, \qquad (SC0)$$

And we have $NAV_0 \geq SCR_0 \Leftrightarrow CurrentSolvencyRatio = \frac{FP_0}{SCR_0} \geq 100\%$.

The required capital can be calculated on the basis of the formula

$$SCR_0 = NAV_0 + K \;/: K = -VaR_{0.5\%}(\delta_1.NAV_1) = -q_{0.5\%}(\delta_1.NAV_1).$$

This formula is true under the assumption that the additional own funds are invested in risk-free assets (the additional capital is capitalized, from $t=0$ on, at the risk free rate). For a more complete analysis of this formula and of the underlying assumptions needed the reader may consult Devineau and Loisel (2009b).

This single-period solvency notion corresponds to the regulatory definition of solvency. The multi-year solvency concept introduced by the ORSA framework leads to several conceptualization issues.

## 2.2. Multi-year solvency

The multi-year framework is particularly efficient to make the undertaking's strategic choices objective. Indeed, the strategic planning time horizon is generally of three to five years. Therefore, the Overall Solvency Needs assessment is particularly relevant for Enterprise Risk Management.

For the company, the major issue is to define its risk limits, coherently with its own strategy planning. In parallel, the firm must define a multi-year solvency constraint in order to assess a required capital level. Basically, a relevant approach requires an in-depth analysis of the questions it is supposed to answer.

A first possibility for this solvency constraint can be to adapt the single-period regulatory constraint. The multi-year solvency notion is then related to owning enough capital today to be able to avoid economic bankruptcy over the whole time horizon with a $p$ threshold. This framework can lead to several implementation issues. One can notice that it is therefore not the type of constraint that has been chosen by most of the French insurance companies.

The adjusted constraint can be framed as

$$\mathbb{P}\left(\bigcap_{t=1}^{T}\{NAV_t \geq 0\}\right) \geq p \Leftrightarrow \mathbb{P}\left(\min_{t\in[|1;T|]}\{NAV_t\} \geq 0\right) \geq p.$$

More generally, it should be possible to consider alternative risk measures (such as Expected Shortfall) or alternative underlying risky variable such as the Solvency Ratio or the Return on Risk Adjusted Capital, as defined in Decupère (2011).

Considering the presented constraint, the ORSA framework raises much thinking about. First, the $p$ probability threshold should typically be chosen rather lower than the 99.5% threshold. Indeed, considering such a constraint enables to develop an overall risk running on both short and long term and in a realistic environment. In this framework it is clearly not appropriate to consider a too restrictive level of solvency. In a broader sense, the $p$ probability threshold can be indexed by the time period in order to relax the constraint over the considered horizon. In addition, the underlying scope considered here is an enlargement of the $SCR$ definition. In particular, as an economic approach, it seems necessary to consider the new business underwritten through the time horizon.

In section 2.3 we formalize different possible approaches of the multi-year solvency constraint. In particular, we introduce a framework that considers the regulatory solvency shortfall probability through the time horizon. In that case the multi-year solvency constraint is based on the Solvency Ratio's chronicle on the whole horizon.

## 2.3. Interpretations of the multi-year solvency

In order to formalize the concepts presented here, we introduce this additional notation. Let $SCR_t$ be the 1-year regulatory capital value at time $t \geq 0$, $\delta_t$ be the discount factor at the end of the first $t \geq 1$ periods, $R_t$ be the profit[1] at time $t \geq 1$, and $T$ be the chosen ORSA time horizon. Additionally, let $\mathcal{F}_t^{RW}$ be the filtration that characterizes the Real-World economic information contained within the $t$ first periods, and $Q_t$ be a Risk-Neutral measure conditioned by the Real-World financial information known at time $t$.

Under this notation, the $NAV$ at the end of the $t^{th}$ period ($t \geq 1$) satisfies the formula

$$NAV_t = \mathbb{E}^{Q_t}\left[\sum_{u \geq 1} \frac{\delta_u}{\delta_t} R_u \,|\, \mathcal{F}_t^{RW}\right].$$

We have (under the same assumption as for $SCR_0$)

$$SCR_t = NAV_t + K \;/\; K = -VaR_{0.5\%}\left(\frac{\delta_{t+1}}{\delta_t}.NAV_{t+1}|\mathcal{F}_t^{RW}\right) = -q_{0.5\%}\left(\frac{\delta_{t+1}}{\delta_t}.NAV_{t+1}|\mathcal{F}_t^{RW}\right).$$

### 2.3.1. Approaches on economic bankruptcy

The approaches on economic bankruptcy aim at looking at whether the undertaking has enough own funds to carry on its business without capital through a chosen time horizon, under the current economic characteristics. Their objective is to translate the undertaking's overall tolerance into a multi-year constraint on the positivity of the $NAV$.

#### 2.3.1.1. Constraint on yearly Net Asset Value distribution

$$\forall t \in [|1;T|], \mathbb{P}(NAV_t \geq 0) \geq p. \qquad (SC1)$$

The aim of such a constraint is to determine the level of own funds at $t = 0$ required to avoid economic bankruptcy at each date of the horizon $[1\,;\,T]$ with the same $p$ probability threshold (with $p$ close to 1).

In a simulation framework, such an approach would lead to the separate analyze of the samples of the random variables $(NAV_t)_{t \in [|1;T|]}$, period after period, without considering path-dependence. At each date, every positive outcome is considered solvent.

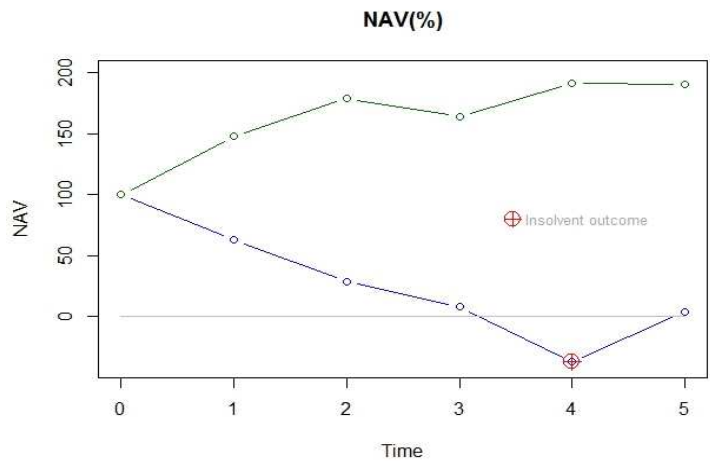

Figure 1: Outcomes of solvent and insolvent NAV through a 5-years horizon (insolvent point targeted in red)

---

[1] Seen as a $NAV$ variation.

Under the assumption that the additional own funds is invested in risk-free assets, the Overall Solvency Needs can be calculated thanks to the formula

$$RequiredCapital_{(SC1)} = NAV_0 + K /: K = -\min_{0<t\leq T}\left(q_{1-p}(\delta_t NAV_t)\right).$$

### 2.3.1.2. Constraint on the Net Asset Value's paths

$$\mathbb{P}\left(\bigcap_{t=1}^{T}\{NAV_t \geq 0\}\right) \geq p. \qquad (SC2)$$

The objective of this constraint is to determine the level of own funds required to avoid economic bankruptcy on the whole horizon $[0\ ;\ T]$ with a $p$ probability threshold.

Considering the intersection $\bigcap_{t=1}^{T}\{NAV_t \geq 0\}$ enables to model the path-dependence of the random process $(NAV_t)_{t\in[|1;T|]}$. In a simulation framework such an approach would lead to consider sample paths of $(NAV_t)_{t\in[|1;T|]}$. Only paths that lead to positive values at each time are considered as solvent.

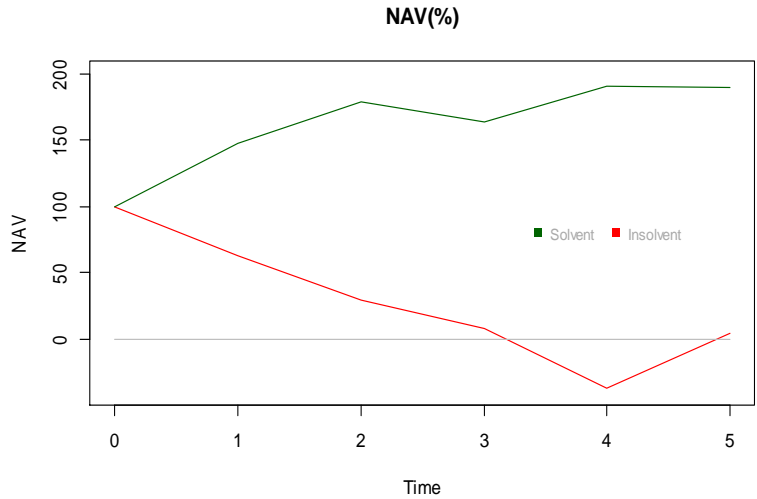

Figure 2: One solvent and one insolvent NAV path for a 5-years horizon

Such a constraint enables to comply with the notion of continuity of business on the whole horizon. Therefore, it seems more realistic from an economic point of view. Under the assumption that the additional own funds are invested in risk-free assets, the Overall Solvency Needs can be calculated thanks to the formula

$$RequiredCapital_{(SC2)} = NAV_0 + K\ /: K = \underset{X}{\text{Argmin}}\left(\mathbb{P}\left(\bigcap_{t=1}^{T}\left\{NAV_t + \frac{X}{\delta_t} \geq 0\right\}\right) = p\right).$$

### 2.3.1.3. Implementation aspects

One possibility of fully simulatory implementation allowing assessment of the Overall Solvency Needs requires the use of multi-year Nested Simulations.

In the Nested Simulations framework, let $Asset_t^n$ be the market value of the firm's asset at time $t$, for the $n^{th}$ primary simulation, $BEL_t^n$ be the Best Estimate of Liabilities of the firm at time $t$, for the $n^{th}$ primary simulation, and $NAV_t^n$ be the $NAV$ of the firm at time $t$, for the $n^{th}$ primary simulation.

The multi-year Nested Simulations procedure can be represented by the following diagram.

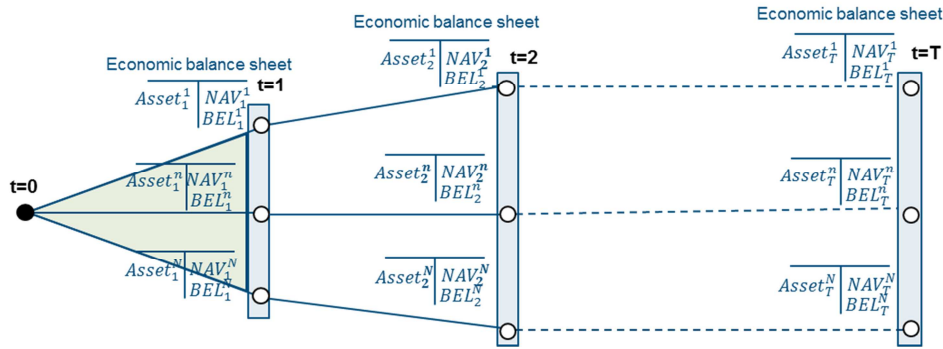

Figure 3: Multi-year Nested Simulations–Generation of sample paths $((NAV_1^n, NAV_2^n, \ldots, NAV_T^n))_{n \in [|1;N|]}$

The primary simulations leading to each node are obtained by Real-World diffusions of the risks. At each node a Risk-Neutral calculation of the firm's asset, of its Best Estimate of Liabilities ($BEL$) and of its $NAV$, is launched.

Such a procedure allows obtaining empirical distributions of $NAV$ paths and percentile values for various thresholds. In practice this approach is hugely time-consuming, especially for life insurance products, for which there is no closed formula to calculate $BEL$ values. However it can be accelerated using efficient proxies of the random variables $(NAV_t)_{t \in [|1;T|]}$. This implementation enables to assess the probability of economic bankruptcy over a chosen time horizon, or the probability of economic bankruptcy at each date of a chosen time horizon, given an own fund's level at $t = 0$. Eventually it permits to calculate an estimate of the Overall Solvency Needs for ($SC1$) and ($SC2$).

Generally speaking, the approaches on economic bankruptcy enable to address the strategic problem of continuity of operations. However, this implementation does not bring any information about the compliance with regulatory solvency requirements on the chosen time horizon. The choice of an approach on solvency shortfalls enables one to deal with this particular issue.

### 2.3.2. Approaches on solvency shortfalls

The approaches on solvency shortfalls translate the undertaking's risk tolerance into a constraint on the value taken by its Solvency Ratio through the time horizon. They aim at assessing whether the undertaking is able to satisfy a regulatory solvency constraint over several years or not.

#### 2.3.2.1. Constraint on yearly Solvency Ratio distributions

$$\forall t \in [|1;T|], \mathbb{P}\left(\frac{NAV_t}{SCR_t} \geq \alpha\right) \geq p. \qquad (SC3)$$

The objective of this constraint is to assess the level of own funds at $t = 0$ required to ensure that the Solvency Ratio stays superior to a level $\alpha$ ($\alpha \geq 0$) at each date of the horizon $[1; T]$ with the same $p$ probability threshold.

In a simulation framework such an approach would lead to analyze separately the outcomes of the random process $\left(\frac{NAV_t}{SCR_t}\right)_{t \in [|1;T|]}$, period after period, without taking path-dependence into account. At each date, every outcome over the limit $\alpha$ is considered as solvent.

The choice of $\alpha$ depends on the risk limits considered in the definition of the firm's risk appetite. If a Solvency Ratio's value is targeted in order to keep a certain credit rating for example, it can be relevant to consider a value $\alpha \geq 100\%$. Indeed such a rating constraint is generally stronger than the regulatory one. However, if the undertaking aims at staying over a lower bound (at least the Minimum Capital Requirement for example), it is possible to consider a value $\alpha \leq 100\%$. Eventually, under the assumption that the additional own funds are invested in risk-free assets, the Overall Solvency Needs can be calculated on the basis of the formula

$$RequiredCapital_{(SC3)} = NAV_0 + K \text{ /: } K = \underset{X}{\text{Argmin}} \left( \underset{0 < t \leq T}{\min} \left( q_{1-p} \left( \frac{NAV_t + \frac{X}{\delta_t}}{SCR_t(X)} \right) \right) \geq \alpha \right).$$

*2.3.2.2. Constraint on the Solvency Ratio's paths*

$$\mathbb{P}\left( \bigcap_{t=1}^{T} \left\{ \frac{NAV_t}{SCR_t} \geq \alpha \right\} \right) \geq p. \qquad (SC4)$$

The objective of this constraint is to assess the level of own funds at $t = 0$ required to ensure the coverage of at least a level $\alpha$ of the regulatory capital on the whole time horizon with a $p$ probability threshold.

In a simulation framework such an approach would lead to focus on each path of the random process $\left( \frac{NAV_t}{SCR_t} \right)_{t \in [|1;T|]}$ as a whole, so as to take path-dependency into account. Only the paths over the $\alpha$ threshold are considered satisfactory.

Under the assumption that the additional own funds are invested in risk-free assets, the Overall Solvency Needs can be calculated on the basis of the formula

$$RequiredCapital_{(SC4)} = NAV_0 + K \text{ /: } K = \underset{X}{\text{Argmin}} \left( \mathbb{P}\left( \bigcap_{t=1}^{T} \left\{ \frac{NAV_t + \frac{X}{\delta_t}}{SCR_t(X)} \geq \alpha \right\} \right) \geq p \right).$$

*2.3.2.3. Implementation aspects*

These constraints can be tested and an estimator of the Overall Solvency Needs can be obtained empirically by using a fully simulatory process (multi-year Nested Simulations). This procedure consists in simulating economic scenarios and calculating a $NAV$ and a 1-year regulatory capital at each node. This approach can be implemented whether the company uses an Internal Model or the Standard Formula approach.

In this framework, let $SCR_t^n$ be the 1-year regulatory capital of the firm at time $t$, for the $n^{th}$ primary simulation.

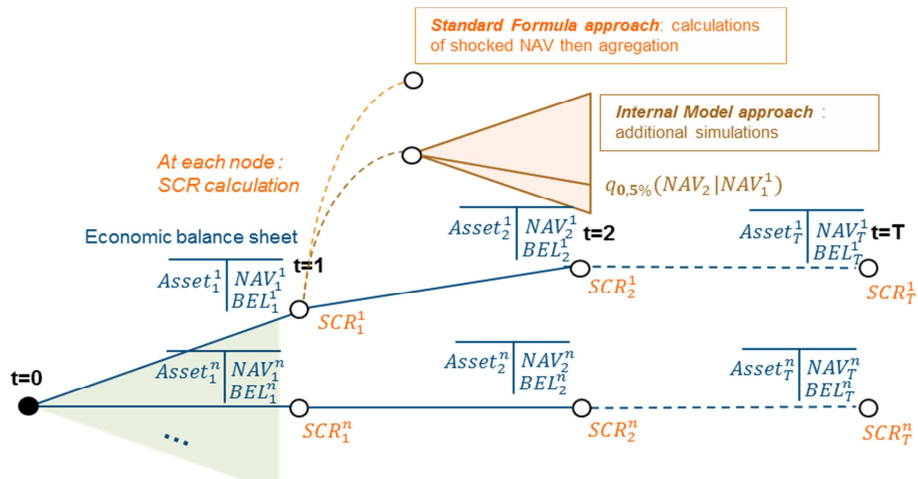

Figure 4: Multi-year Nested Simulations –Generation of joint sample paths $\left((NAV_1^n, NAV_2^n, \ldots, NAV_T^n)\right)_{n \in [|1;N|]}$ and $\left((SCR_1^n, SCR_2^n, \ldots, SCR_T^n)\right)_{n \in [|1;N|]}$

In practice a fully simulatory approach is particularly difficult to implement but can be made extremely less time-consuming using proxies and the Standard Formula approach to obtain the $SCR$ outcomes. This implementation enables to assess the probability of meeting a minimum level $\alpha$ of the regulatory capital on a chosen time horizon, or the probability of meeting a minimum level $\alpha$ of the regulatory capital at each date of a chosen time horizon, given an own funds level at $t = 0$. It eventually enables the calculation of an estimate of the Overall Solvency Needs as expressed in $(SC3)$ and $(SC4)$.

### 2.3.3. Approaches on multi-deterministic scenarios

Both approaches presented above lead to major implementation issues. Therefore, the insurance industry tends towards using of a more pragmatic approach, introducing a looser solvency constraint. The implementation of an approach on multi-deterministic scenarios consists in testing a solvency constraint on a limited number of multi-year stressed scenarios. In practice these scenarios must be chosen depending on their adversity level and their relevance to the underlying risks. Once a set of $J$ stressed scenarios is calibrated, the constraint can be stated as

$$\forall j \in [|1;J|], \forall t \in [|1;T|], \frac{NAV_t^j}{SCR_t^j} \geq \alpha. \qquad (SC5)$$

The procedure to test the compliance with this constraint is easier to implement. Indeed, it is possible to assess the outcomes $\left(NAV_t^j, SCR_t^j\right)_{j,t}$ and to test the solvency constraint based on this small number of outcomes.

The major issue of these approaches is the choice of the stressed scenarios. Basically these scenarios are marginally or jointly stressed economic scenarios. The underlying risk factors must be chosen for their significance and the shock intensity must be carefully defined. In particular, for better economic realism, it can be relevant to integrate countercyclicality effects, mean reversion mechanisms and management actions. In practice a small number of scenarios are considered and the Standard Formula approach is chosen for the $SCR$ calculations. Therefore few calculations are necessary to test the constraint.

The aim of these approaches is to assess whether the undertaking is able to meet a minimum level $\alpha$ of the regulatory capital through a chosen time horizon and for the selected stressed scenarios.

Under the assumption that the additional own funds are invested in risk-free assets, the Overall Solvency Needs can be calculated on the basis of the formula

$$RequiredCapital_{(SC5)} = FP_0 + K \ /: K = -\min_{\substack{0<t\leq T \\ 1\leq j\leq J}} \left(\delta_t^j \left(FP_t^j - \alpha \times SCR_t^j\right)\right).$$

It should be theoretically necessary to take the impact of the initial capital addition on the $SCR$, $K$, into account, for each period. However, these pragmatic approaches usually assume that this impact is negligible.

Eventually, the operational relevance of such an approach is obvious but it leads to difficulties in the choice of the stressed scenarios. Moreover, they provide less information compared to the previous approaches.

In Section 3, we develop quantitative proxy methodologies that provide satisfactory alternatives to fully Nested Simulations procedures, for life insurance economic balance sheet projections. These methodologies allow to obtain joint sample paths of the $NAV$ and of the $SCR$.

# 3. Quantitative methodologies for the Overall Solvency Needs assessment in a life insurance company framework

Section 3 presents in a formal way the fully simulatory implementation and its proxy alternatives. Their common goal is to enable the use of an approach on solvency shortfalls in order to assess the Overall Solvency Needs associated to a life insurance company. The study will be carried out in a context where the $SCR$ is calculated through the Standard Formula approach.

## 3.1. Multi-year Nested Simulations

### 3.1.1. Obtaining of a multi-year Net Asset Value distribution

We aim at obtaining empirical outcomes of the variables $(NAV_t)_{t\in[|1;T|]}$, satisfying the following equation,

$$\forall t \in [|1;T|], NAV_t = \mathbb{E}^{Q_t}\left[\sum_{u\geq 1}\frac{\delta_u}{\delta_t}R_u \,|\mathcal{F}_t^{RW}\right] = \sum_{u=1}^{t}\frac{\delta_u}{\delta_t}R_u + \mathbb{E}^{Q_t}\left[\sum_{u>t}\frac{\delta_u}{\delta_t}R_u \,|\mathcal{F}_t^{RW}\right].$$

The Nested Simulations procedure is an empirical methodology based on a Monte-Carlo valuation of $\mathbb{E}^{Q_t}\left[\sum_{u>t}\frac{\delta_u}{\delta_t}R_u\,|\mathcal{F}_t^{RW}\right]$. In practice, the implementation of this methodology in order to obtain $N$ outcomes $(NAV_1^n, \dots NAV_T^n)_{n\in[|1;N|]}$ follows this sequence:

Step 1: Real-World diffusion of $N$ random primary economic scenarios through the period $[0\,;T]$.

Step 2: For each primary scenario $n$ and date $t \in [\![1;T]\!]$:

- Use of a simulation model to calculate the outcomes $\left(\frac{\delta_u^n}{\delta_t^n}R_u^n\right)_{u\in[|1;t|]}$, the profit values at each date through $[0\,;t]$, capitalized until the $t^{th}$ period and conditioned by scenario $n$.
- Diffusion of $P$ Risk-Neutral secondary economic scenarios through the period $[t\,;t+H]^2$, conditioned by the Real-World economic outcomes of scenario $n$ at time $t$. These scenarios must be i.i.d. conditionally to $\mathcal{F}_t^{RW}$.
- Calculation of the related values $\left(\frac{\delta_u^{n,p}}{\delta_t^n}R_u^{n,p}\right)_{u\in[|t+1;t+H|]}$, for each secondary scenario $p$,.
- Construction of an empirical estimator of the value $NAV_t^n = \sum_{u=1}^{t}\frac{\delta_u^n}{\delta_t^n}R_u^n + \mathbb{E}^{Q_t}\left[\sum_{u=t+1}^{t+H}\frac{\delta_u}{\delta_t}R_u\,|\mathcal{F}_t^n\right]$, $\mathcal{F}_t^n$ being the filtration that characterizes the Real-World economic information contained within the primary simulation $n$,

$$\widehat{NAV}_t^n = \sum_{u=1}^{t}\frac{\delta_u^n}{\delta_t^n}R_u + \frac{1}{P}\sum_{p=1}^{P}\sum_{u=t+1}^{t+H}\frac{\delta_u^{n,p}}{\delta_t^n}R_u^{n,p}.$$

Eventually, the Nested Simulations procedure requires two kinds of stochastic economic scenarios tables. One table of $N$ primary scenarios generated between $t=0$ and $t=T$ under the historical probability. Then, for each primary scenario and each date $t \in [|1;T|]$, one conditional table of $P$ secondary scenarios generated through the period $[t\,;t+H]$, under the probability measure $Q_t$.

---

[2] $H$ is the liability extinction horizon (ranging operationally from 30 to 50 years).

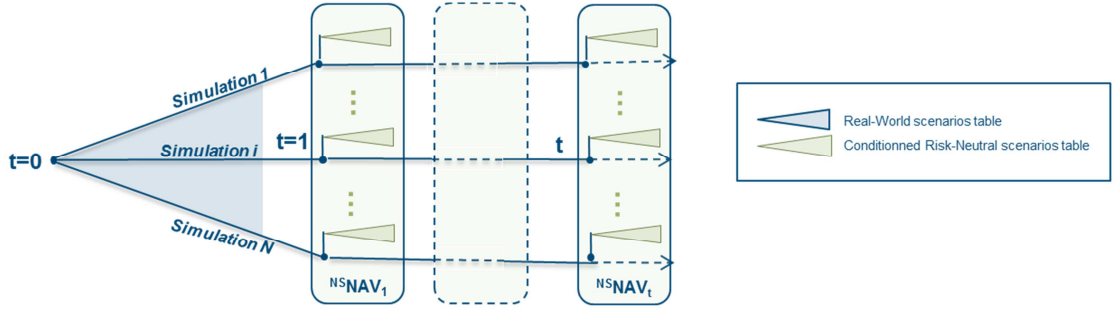

Figure 5:Multi-year Nested Simulations to obtain Net Asset Value distributions

This methodology enables to obtain empirical approximations, assumed exact operationally as soon as $P$ is large enough, of the random vector $((NAV_1^n, NAV_2^n, \ldots, NAV_T^n))_{n\in[|1;N|]}$. By definition we have, $\forall n \in [|1;N|]$,

$$\widehat{NAV}_t^n \xrightarrow[P\to+\infty]{a.s.} NAV_t^n \text{ (strong law of large numbers)}.$$

In the rest of the paper we will denote by $\widehat{NAV}_t$ the random variable associated to the approximated $NAV$ obtained by using a Nested Simulations methodology.

### 3.1.2. Implementation to obtain a multi-year Solvency Capital Requirement distribution

Taking the Standard Formula approach into account, it is possible to derive the sample paths of $SCR$ by duplicating this procedure. Indeed, the application of marginal shocks to the economic conditions at each node, before conditioning the Risk-Neutral tables, enables to calculate marginally shocked $NAV$.

Say we consider $I$ different risks and let $\mathcal{F}_{t^+}^{n,risk(i)}$ be the filtration that characterizes the Real-World economic information contained within the primary simulation $n$ until date $t$ and the marginally shocked economic conditions associated to the risk number $i \in [\![1;I]\!]$ instantaneously applied after the date $t$. We have, for each primary scenario $n \in [\![1;N]\!]$ and $i \in [\![1;I]\!]$,

$$NAV_t^{n,risk(i)} = \sum_{u=1}^{t} \frac{\delta_u^n}{\delta_t^n} R_u + \mathbb{E}^{Q_t}\left[\sum_{u>t} \frac{\delta_u}{\delta_t} R_u \Big| \mathcal{F}_{t^+}^{n,risk(i)}\right] \approx \sum_{u=1}^{t} \frac{\delta_u^n}{\delta_t^n} R_u + \mathbb{E}^{Q_t}\left[\sum_{u=t+1}^{t+H} \frac{\delta_u}{\delta_t} R_u \Big| \mathcal{F}_{t^+}^{n,risk(i)}\right],$$

with $NAV_t^{n,risk(i)}$ being the $NAV$ at time $t$ associated to the primary scenario $n$ and shocked using the Standard Formula risk $i$ (equity, interest rates, lapse,…). We obtain empirical estimators of the following shape

$$\widehat{NAV}_t^{n,risk(i)} = \sum_{u=1}^{t} \frac{\delta_u^n}{\delta_t^n} R_u^n + \frac{1}{P}\sum_{p=1}^{P}\sum_{u=t+1}^{t+H} \frac{\delta_u^{n,p\,risk(i)}}{\delta_t^n} R_u^{n,p\,risk(i)}.$$

Using the Standard Formula aggregation matrices it is now possible to obtain the joint sample paths $((\widehat{NAV}_1^n, \ldots, \widehat{NAV}_T^n))_{n\in[|1;N|]}$ and $((\widehat{SCR}_1^n, \ldots, \widehat{SCR}_T^n))_{n\in[|1;N|]}$. We detail below the Standard Formula calculations needed to obtain the sample paths $((\widehat{SCR}_1^n, \ldots, \widehat{SCR}_T^n))_{n\in[|1;N|]}$.

Let $M$ be the set of risk modules considered, $Risk_m$ be the set of risks in module $m \in [\![1;M]\!]$. $Risk_m \subset [\![1;I]\!]$, $\hat{C}_t^{n,risk\,i}$ be the stand-alone capital associated to risk $i$ at time $t$ and for the $n^{th}$ scenario, obtained by using Nested Simulations, and $\widehat{SCR}_{m,t}^n$ be the $SCR$ associated to module $m$ at time $t$ and for the $n^{th}$ scenario, obtained by using Nested Simulations. Moreover, let $\left(\rho_m^{i,j}\right)_{i,j\in R_m^2}$ be the QIS correlation matrix to aggregate the capitals associated to risk module $m$, and $(\rho^{k,l})_{k,l\in[\![1;M]\!]^2}$ be the QIS correlation matrix to aggregate the $SCR$ associated to the various risk modules.

The $SCR$ outcomes are calculated on the basis of the following formulas.

$\forall t \in [\![1;T]\!], \forall n \in [\![1;N]\!]$, risk capitals calculation,

$$\forall i \in [\![1;I]\!], \hat{C}_t^{n,risk\ i} = \max\{\widehat{NAV}_t^n - \widehat{NAV}_t^{n,riski}; 0\}.$$

Intra-modular aggregation,

$$\forall m \in [\![1;M]\!], \widehat{SCR}_{m,t}^n = \sqrt{\sum_{i,j \in Risk_m^2} \rho_m^{i,j} \cdot \hat{C}_t^{n,risk\ i} \cdot \hat{C}_t^{n,risk\ j}}.$$

Inter-modular aggregation[3],

$$\widehat{SCR}_t^n = \sqrt{\sum_{k,l \in [\![1;M]\!]^2} \rho^{k,l} \cdot \widehat{SCR}_{k,t}^n \cdot \widehat{SCR}_{l,t}^n}.$$

There is no closed formula to calculate the $NAV$ and the $SCR$ except when relatively strong model assumptions are made, as in Bonnin and al. (2012). This Monte-Carlo methodology is the most precise approach for deriving joint sample paths of the $NAV$ and $SCR$. However it is extremely time-consuming due to the large number of simulations needed to obtain efficient estimators.

The various alternative approaches developed in Subsections 3.2 and 3.3 aim at approximating the outcomes obtained by implementing such a procedure. The use of these proxies allows for a significant acceleration of this fully simulatory framework.

### 3.2. Multi-year Curve Fitting

The aim of a Curve Fitting approach is to calibrate a polynomial function replicating either the Best Estimate of Liabilities or the $NAV$. Various kinds of regressors can be used but they are easy to calculate and to handle (in this article we will consider polynomial functions of the elementary risk factors of the primary simulations). For more insight about the elementary risk factors definition, the reader may consult Devineau and Chauvigny (2010). This approach is already used in a single-period framework in order to replicate the $NAV$ at time $t = 1$ and assess the economic capital. It can be adapted to a larger time horizon with satisfactory results.

#### 3.2.1. Formalization to obtain a multi-year Net Asset Value distribution

Here, we consider a polynomial proxy that replicates the $NAV$ in a straightforward fashion. For the sake of simplicity we will only consider two risks: the risk on the stock index and on interest rates.[4]

The gist of the method is the following. The $NAV$ at time $t$ depends on the economic information through the period $[0; t]$, $NAV_t = \mathbb{E}^{Q_t}\left[\sum_{u=1}^{t+H} \frac{\delta_u}{\delta_t} R_u \mid \mathcal{F}_t^{RW}\right]$.

When related to the $n^{th}$ outcome obtained by implementing a Nested Simulations methodology, we use the approximated $NAV$ at date $t$ introduced above,

---
[3] Here the Solvency Capital Requirement is identified to the Basic Solvency Capital Requirement.
[4] We only focus in this article on the level risk and do not consider the volatility risk.

$$\widehat{NAV}_t^n = \sum_{u=1}^{t} \frac{\delta_u^n}{\delta_t^n} R_u^n + \frac{1}{P} \sum_{p=1}^{P} \sum_{u=t+1}^{t+H} \frac{\delta_u^{n,p}}{\delta_t^n} R_u^{n,p}.$$

Eventually, every primary economic scenario can be synthesized into sets of yearly elementary risk factors with little or no loss of information, denoted ($^s\varepsilon_1^n$, $^s\varepsilon_2^n$, ..., $^s\varepsilon_t^n$) (resp. ($^{ZC}\varepsilon_1^n$, $^{ZC}\varepsilon_2^n$, ..., $^{ZC}\varepsilon_t^n$)) for the stock (resp. interest rate) yearly elementary risk factors. Where $^s\varepsilon_u^n$ (resp. $^{ZC}\varepsilon_u^n$) stands for the stock (resp. interest rate) elementary risk factor related to the simulation $n$ and the period $[u-1, u]$.

These elementary risk factors can be extracted from the primary scenarios table considering formulas similar to those proposed in Devineau and Chauvigny (2010).

We have: $NAV_t^n \approx f_t(\,^s\varepsilon_1^n, \,^{ZC}\varepsilon_1^n, \,^s\varepsilon_2^n, \,^{ZC}\varepsilon_2^n, ..., \,^s\varepsilon_t^n, \,^{ZC}\varepsilon_t^n)$

for a certain polynomial function $f_t$.

The notation $Z^{tr}$ stands further for the transpose of a matrix or vector $Z$.

The implementation of this methodology follows this sequence for each date $t \in [\![1; T]\!]$:

Step 1: Calculation of a small number $N'$ of outcomes $\left(\widehat{NAV}_t^{n'}\right)_{n' \in [\![1; N']\!]}$.

Step 2: Calibration of the optimal set of regressors[5] $X_t = (Intercept, \,^1X_t, ..., \,^kX_t)$ [6], with $^iX_t = \,^s\varepsilon_t^{x_i} \cdot \,^{ZC}\varepsilon_t^{y_i}$, for each $i \in [\![1; k]\!]$ and certain positive integers $x_i$ and $y_i$. Then determination of $\hat{\beta}_t = (\,^I\hat{\beta}_t, \,^1\hat{\beta}_t, ..., \,^k\hat{\beta}_t)^{tr}$, the ordinary least squares (OLS) parameters estimator of the multiple regression $\widehat{NAV}_t = X_t. \beta_t + u_t$ where $\widehat{NAV}_t$ is approximated by its conditional expectation given the $\sigma$-field generated by the regressors $X_t$, considered as a linear combination of the regressors. For more insight about multiple regression models the reader may consult Saporta (2011).

The underlying assumption of this model can therefore be written

$$\exists \beta_t, \mathbb{E}[\widehat{NAV}_t | X_t] = X_t. \beta_t.$$

Step 3: Generation of $N$ independent outcomes of the optimal regressors $\left(X_t^n = (1, \,^1X_t^n, ..., \,^kX_t^n)\right)_{n \in [\![1; N]\!]}$ and calculation of the approximated distribution $\left(^{CF}NAV_t^n = X_t^n. \hat{\beta}_t\right)_{n \in [\![1; N]\!]}$.

In order to formalize the methodology, we introduce the following notation. Let $^{\widehat{NAV}}Y_t = \left(\widehat{NAV}_t^1, ..., \widehat{NAV}_t^{N'}\right)^{tr}$ be the Nested Simulations outcomes used in the calibration step, and $R(X_t) = \begin{pmatrix} X_t^1 \\ \vdots \\ X_t^{N'} \end{pmatrix}$ be the matrix of the $N'$ outcomes of the optimal set of regressors $X_t$, related to the outcomes $\left(\widehat{NAV}_t^{n'}\right)_{n' \in [\![1; N']\!]}$ used in the calibration step.

With this notation, the parameters estimator is obtained by solving the optimization program

$$\hat{\beta}_t = \underset{\beta_t}{\mathrm{Argmin}} \left\{ \left\| ^{\widehat{NAV}}Y_t - R(X_t). \beta_t \right\|_1^2 \right\}.$$

---

[5] For more insight concerning the calibration of the optimal set of regressors see Subsection 3.2.2.
[6] Here, $k$ is the number of regressors (not including the intercept) in the calibrated optimal set of regressors. In all generality $k$ depends of time $t$. This indexation is omitted for the sake of simplicity.

Under the assumption that the matrix $R(X_t)^{tr}.R(X_t)$ is invertible (satisfied in practice for $k$ small enough) this program has a unique solution,

$$\hat{\beta}_t = \left(R(X_t)^{tr}.R(X_t)\right)^{-1} R(X_t)^{tr}.\widehat{NAV}Y_t.$$

### 3.2.2. Implementation issues

This implementation shows two major issues: the choice of the $N'$ primary scenarios used if the calibration step and the choice of the set of optimal regressors.

The scenarios considered so as to calibrate the polynomial function must enable to obtain an estimator $\hat{\beta}_t$ close to the optimal $\beta_t$ in order to efficiently replicate the overall distribution of $NAV_t$. Two approaches are usually considered to achieve this. First it can be interesting to consider well-dispersed primary scenarios. It is also possible to consider "extreme" primary scenarios, according to a metric of the scenario's adversity threshold, for example a chosen norm on the underlying risk factor, as developed in Devineau and Loisel (2009).

Both approaches can lead to satisfactory empirical results, depending on the characteristics of the life insurance products, the economic conditions, etc.

Considering the second issue, the general procedure aims at maximizing the $R^2$ of the considered regression under a significance constraint on the chosen covariates. Various automatic approaches can be used with satisfactory results such as the Stepwise Regression procedures. For more developments about these approaches see Draper and Smith (1981).

### 3.2.3. Formalization to obtain a multi-year Solvency Capital Requirement distribution

In order to obtain a multi-year $SCR$ distribution joint with the sample paths $\left({}^{CF}NAV_1^n, {}^{CF}NAV_2^n, \ldots, {}^{CF}NAV_T^n\right)_{n\in[1,N]}$, it is possible to duplicate this procedure and calibrate proxies "after shock" : $X_t^{risk(i)}.\hat{\beta}_t^{risk(i)}$, $X_t^{risk(i)}$ being the optimal set of regressors of the proxy "after shock i" (equity or interest rate up/down) and $\hat{\beta}_t^{risk(i)}$ the associated parameters estimator.

Basically, it is necessary to duplicate fully the procedure in order to calibrate new proxies "after shock", to obtain approximated marginally shocked $NAV$. Indeed, the Standard Formula shocks must be applied after the rebalancing of assets and the asset-mix dependent "central" proxies are not adapted to this new framework. Therefore they cannot be used to approximate the marginally shocked $NAV$.

The proxies "after shock" are obtained by considering a calibration step upon $N'$ shocked outcomes $\left(\widehat{NAV}_t^{n',risk\,(i)}\right)_{n'\in[\![1;N']\!]}$, for each considered Standard Formula risk $i$ and date $t\in[\![1;T]\!]$. Then the Standard Formula aggregation of the central and shocked outcomes, $\left(X_t^n.\hat{\beta}_t\right)_{n\in[|1;N|]}$ and $\left(X_t^{n,risk(i)}.\hat{\beta}_t^{riski}\right)_{n\in[|1;N|]}$, enables to obtain joint sample paths of the $NAV$ and $SCR$. One can notice that, in practice, the optimal sets $X_t^{risk(i)}$, for all $i$ (equity or interest rate up/down), are very close to $X_t^n$.[7]

---

[7] In practice, it seems even possible not to calibrate specific sets of regressors for the proxies "after shock" and just to use the optimal set of regressors calibrated for the "central" proxies. This process has been used in Section 4. It accelerates the methodology and provides satisfactory results.

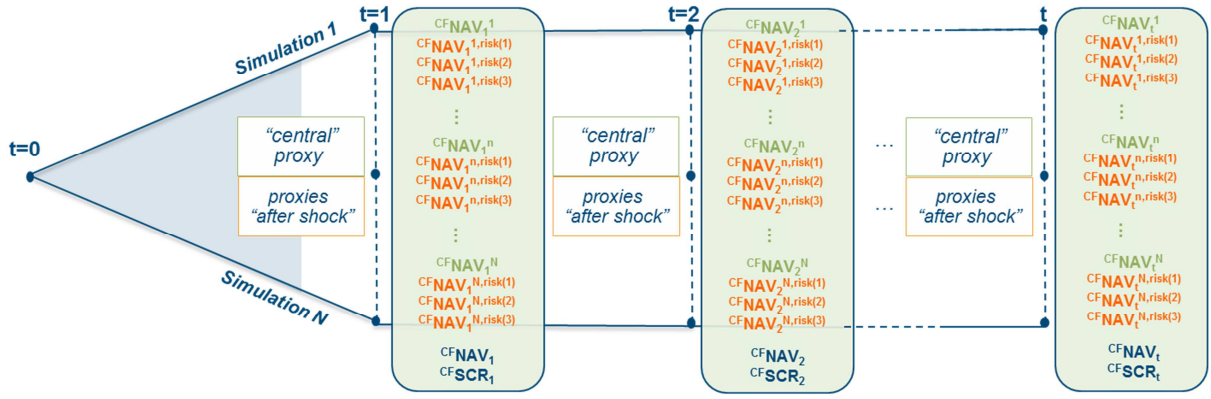

Figure 6: Curve Fitting approach - Simulation of approximated jointsample paths of NAV and SCR (3 shocks)

$N'$ being usually quite small, this alternative implementation greatly accelerates the Nested Simulations procedure with satisfactory results.

### 3.3. Multi-year Least Squares Monte-Carlo

It is generally considered that the so-called Least Squares Monte-Carlo ($LSMC$) methodology has first been introduced in Longstaff and Schwartz (2001) as a methodology for approximating the value of American options. However, the implementation of the Least Squares Monte-Carlo in the insurance framework is quite different of the financial valuation algorithm developed by Longstaff and Schwartz.

American options pricing is particularly tricky and their valuation leads to a step by step maximization problem. Various methodologies have been proposed to work around this valuation issue. Longstaff and Schwarz consider the framework of an American option that can only be exercised at discrete times $t_1, t_2, \dots, t_N$. They propose a backward inducted methodology that uses polynomial functions of the underlying asset to approximate the conditionally expected value of the Net Present Value of future cash-flows at each exercise date. At each date $t_k$ the conditionally expected value is compared to the immediate exercise value in order to identify the exercise decision. Then the process is duplicated at time $t_{k-1}$ and recursively until the initial valuation time.

In the insurance framework, the $LSMC$ is a forward-looking approach used to approximate the $NAV$ or the Best Estimate of Liabilities values by using polynomial functions. Basically this methodology is very close to Curve Fitting. The major difference between both proxy approaches lies in the calibration step.

#### 3.3.1. Formalization to obtain a multi-year distribution of Net Asset Value

This alternative approach introduces the notion of $NPV_t$, a Net Present Value of margins at each date $t \in [\![1;T]\!]$, adding with the capitalized profits through$[0\,;t]$, defined as

$$NPV_t = \sum_{u \geq 1} \frac{\delta_u}{\delta_t} R_u = \sum_{u=1}^{t+H} \frac{\delta_u}{\delta_t} R_u.$$

One can decompose the $\widehat{NAV}_t$ outcomes as a sum of $NPV_t$ outcomes:

$$\widehat{NAV}_t^n = \sum_{u=1}^{t} \frac{\delta_u^n}{\delta_t^n} R_u^n + \frac{1}{P} \sum_{p=1}^{P} \sum_{u=t+1}^{t+H} \frac{\delta_u^{n,p}}{\delta_t^n} R_u^{n,p} = \frac{1}{P} \sum_{p=1}^{P} NPV_t^{n,p}.$$

With introduction of $NPV_t^{n,p}$, the Net Present Value of margins at time $t$ associated to the $n^{th}$ primary scenario and the $p^{th}$ secondary scenario, adding with the capitalized profits through $[0\,;t]$.

The basement of the methodology, adapted to our insurance framework, comes from the idea that the use of $\widehat{NAV}_t$ outcomes for the calibration of the polynomial proxies introduces redundancy of information that can be avoided by considering $NPV_t$ outcomes which calculation relies on a single simulated path.

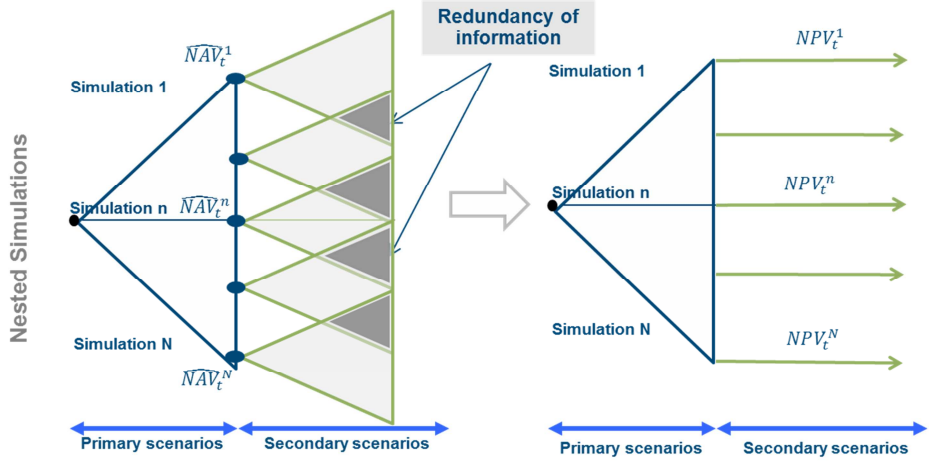

Figure 7: Illustration of the redundancy of information introduced by Nested Simulations

The implementation of this methodology follows this sequence for each date $t \in [\![1; T]\!]$:

Step 1: Calculation of a large number $N'$ of outcomes $\left( NPV_t^{n',1} = NPV_t^{n'} \right)_{n' \in [\![1; N']\!]}$ obtained by considering independent secondary simulations. Basically these outcomes can be considered as a sample of $\widehat{NAV}_t$ obtained by considering one secondary scenario ($P = 1$) per primary scenario. All the secondary scenarios must be generated independently.

Step 2: Calibration of the optimal set of regressors $X_t = \left( Intercept, {}^1X_t, \ldots, {}^kX_t \right)$, with ${}^iX_t = {}^s\varepsilon_t^{x_i} . {}^{ZC}\varepsilon_t^{y_i}$, for each $i \in [\![1; k]\!]$ and certain positive integers $x_i$ and $y_i$. Then determination of $\hat{\beta}_t = \left( {}^I\hat{\beta}_t, {}^1\hat{\beta}_t, \ldots, {}^k\hat{\beta}_t \right)^{tr}$, the OLS parameters estimator of the multiple regression $NPV_t = X_t . \beta_t + v_t$ where $NPV_t$ is approximated by its conditional expectation given the $\sigma$-field generated by the regressors $X_t$, considered as a linear combination of the regressors.

The underlying assumption of this model can therefore be written
$$\exists \beta_t, \mathbb{E}[NPV_t | X_t] = X_t . \beta_t.$$

Step 3: Generation of $N$ independent outcomes of the optimal regressors $\left( X_t^n = (1, {}^1X_t^n, \ldots, {}^kX_t^n) \right)_{n \in [\![1; N]\!]}$ and calculation of an approximated distribution of $NAV_t$, and not only of $NPV_t$, $\left( {}^{LSMC}NAV_t^n = X_t^n . \hat{\beta}_t \right)_{n \in [\![1; N]\!]}$. Indeed, as proven in Subsection 4.1.3, the $LSMC$ and the Curve Fitting approaches both converge asymptotically towards the same $NAV_t$ polynomial proxy.

A calculation of $NPV_t$ is much quicker than a calculation of $\widehat{NAV}_t$ as considered in the Curve Fitting methodology (with $P \gg 1$). However it brings little information on the value of the related $NAV_t$. This justifies the need to consider a large $N'$.

In order to formalize the methodology, we introduce the following notation. Let ${}^{NPV}Y_t = \left( NPV_t^1, \ldots, NPV_t^{N'} \right)^{tr}$ be the outcomes used in the calibration step, and $R(X_t) = \begin{pmatrix} X_t^1 \\ \vdots \\ X_t^{N'} \end{pmatrix}$ be the matrix of the $N'$ outcomes of the optimal set of regressors $X_t$ related to the sample $\left( NPV_t^{n'} \right)_{n' \in [\![1; N']\!]}$ used in the calibration step.

Under this notation, the parameters estimator is obtained by solving the optimization program

$$\hat{\beta}_t = \underset{\beta_t}{\text{Argmin}} \left\{ \left\| {}^{NPV}Y_t - R(X_t).\beta_t \right\|_1^2 \right\}.$$

Under the assumption that the matrix $R(X_t)^{tr}.R(X_t)$ is invertible (satisfied in practice for $k$ small enough) this program has a unique solution,

$$\hat{\beta}_t = \left( R(X_t)^{tr}.R(X_t) \right)^{-1} R(X_t)^{tr}. {}^{NPV}Y_t.$$

### 3.3.2. Implementation issues and formalization to obtain a multi-year distribution of Solvency Capital Requirement

The implementation issues raised by this procedure are similar to those already considered in the Curve Fitting methodology. Similarly one can duplicate the procedure in order to obtain shocked proxies, thus enabling to calculate approximated joint sample paths of shocked $NAV$ through the chosen time horizon. These $LSMC$ outcomes can be aggregated using the Standard Formula matrices to obtain joint outcomes of the $NAV$ and of the $SCR$.

## 4. Theoretical formalization and comparison of the polynomial approaches

The main objectives of Section 4 are to prove the asymptotic convergence of both Curve Fitting and $LSMC$ estimators and to establish a result enabling the comparison of both Curve Fitting and $LSMC$ efficiency. A framework leading to formulas for the estimator error factor in both Curve Fitting and Least Squares Monte-Carlo approaches has been investigated in Kalberer (2012).

In order to theoretically formalize both methodologies we introduce the following notation. Let $\mathbb{P}$ be the Real-World probability measure, and $(\mathbb{P} \otimes \mathcal{Q})_t$ be the probability measure under which our variables indexed by time $t$ are simulated. The indexation by $(\mathbb{P} \otimes \mathcal{Q})_t$ will be omitted for the sake of simplicity.

Recall the definition of the following random variables,

$$NPV_t = \sum_{u \geq 1} \frac{\delta_u}{\delta_t} R_u,$$

and

$$\widehat{NAV}_t = \frac{1}{P} \sum_{p=1}^{P} NPV_t^p,$$

Considering $P$ i.i.d. random variables conditionally to $\mathcal{F}_t^{RW}$, that have the same law as $NPV_t$,

$$NPV_t^1, NPV_t^2, \dots, NPV_t^P.$$

We have, by definition of $NAV_t$ ,

$$NAV_t = \mathbb{E}[NPV_t \,|\mathcal{F}_t^{RW}] = \mathbb{E}[\widehat{NAV}_t | \mathcal{F}_t^{RW}].$$

### 4.1. Preliminary results

#### 4.1.1. Curve Fitting formalization

The model considered here is the following

$$\widehat{NAV}_t = X_t \cdot {}^{(1)}\beta_t + u_t, \quad (1)$$

where $\widehat{NAV}_t$ is approximated by its conditional expectation given the $\sigma$-field generated by the regressors $X_t$, considered as a linear combination of the regressors.

The underlying assumption of this model can therefore be written[8]

$$\exists\, {}^{(1)}\beta_t, \mathbb{E}[\widehat{NAV}_t | X_t] = X_t \cdot {}^{(1)}\beta_t.$$

We introduce the following notation

- $\mathbb{V}[u_t] = {}^u\sigma_t^2$,
- $\mathbb{E}[\mathbb{V}[NPV_t | \mathcal{F}_t^{RW}]] = {}^{NPV}\sigma_t^2$,
- $\mathbb{V}[NAV_t] = {}^{NAV}\sigma_t^2$.

Under this notation, we have

$$^{NAV}\sigma_t^2 + \frac{{}^{NPV}\sigma_t^2}{P} = \mathbb{V}[X_t \cdot {}^{(1)}\beta_t] + {}^u\sigma_t^2. \quad (2)$$

Indeed,

$$\mathbb{V}[\widehat{NAV}_t] = \mathbb{V}\left[\mathbb{E}[\widehat{NAV}_t | \mathcal{F}_t^{RW}]\right] + \mathbb{E}\left[\mathbb{V}[\widehat{NAV}_t | \mathcal{F}_t^{RW}]\right] = \mathbb{V}[NAV_t] + \mathbb{E}\left[\mathbb{V}[\widehat{NAV}_t | \mathcal{F}_t^{RW}]\right]$$

$$\mathbb{V}[\widehat{NAV}_t] = {}^{NAV}\sigma_t^2 + \frac{1}{P}\mathbb{E}[\mathbb{V}[NPV_t | \mathcal{F}_t^{RW}]] = {}^{NAV}\sigma_t^2 + \frac{{}^{NPV}\sigma_t^2}{P},$$

Moreover,

$$\mathbb{E}[\widehat{NAV}_t | X_t] = X_t \cdot {}^{(1)}\beta_t$$

implies that

$$\mathbb{E}[X_t \cdot u_t] = \mathbb{E}\left[\mathbb{E}[X_t \cdot (\widehat{NAV}_t - X_t \cdot {}^{(1)}\beta_t) | \mathcal{F}_t^{RW}]\right] = (0, \dots, 0).$$

We have therefore,

$$\mathbb{V}[\widehat{NAV}_t] = \mathbb{V}[X_t \cdot {}^{(1)}\beta_t + u_t] = \mathbb{V}[X_t \cdot {}^{(1)}\beta_t] + \mathbb{V}[u_t]$$

$$\mathbb{V}[\widehat{NAV}_t] = \mathbb{V}[X_t \cdot {}^{(1)}\beta_t] + {}^u\sigma_t^2.$$

#### 4.1.2. Least Squares Monte-Carlo formalization

The model considered here is the following

$$NPV_t = X_t \cdot {}^{(2)}\beta_t + v_t. \quad (3)$$

---

[8] This assumption can be loosened by considering $\mathbb{E}[X_t \cdot u_t] = 0$. This leads to the same results.

In this model, $NPV_t$ is approximated by its conditional expectation given the $\sigma$-field generated by the regressors $X_t$, considered as a linear combination of the regressors.

The underlying assumption of this model can therefore be written as

$$\exists\, ^{(2)}\beta_t, \mathbb{E}[NPV_t|X_t] = X_t \cdot {}^{(2)}\beta_t.$$

We introduce the following notation,

$$\mathbb{V}[v_t] = {}^v\sigma_t^2.$$

Under this notation, we have

$$^{NAV}\sigma_t^2 + {}^{NPV}\sigma_t^2 = \mathbb{V}[X_t^{n'} \cdot {}^{(2)}\beta_t] + {}^v\sigma_t^2. \quad (4)$$

Indeed,

$$\mathbb{V}[NPV_t] = \mathbb{V}\big[\mathbb{E}[NPV_t|\mathcal{F}_t^{RW}]\big] + \mathbb{E}\big[\mathbb{V}[NPV_t|\mathcal{F}_t^{RW}]\big] = \mathbb{V}[NAV_t] + \mathbb{E}\big[\mathbb{V}[NPV_t|\mathcal{F}_t^{RW}]\big]$$

$$\mathbb{V}[NPV_t] = {}^{NAV}\sigma_t^2 + {}^{NPV}\sigma_t^2,$$

Moreover,

$$\mathbb{E}[NPV_t|X_t] = X_t \cdot {}^{(2)}\beta_t$$

implies that

$$\mathbb{E}[X_t \cdot v_t] = \mathbb{E}\Big[\mathbb{E}\big[X_t \cdot (NPV_t - X_t \cdot {}^{(2)}\beta_t)|\mathcal{F}_t^{RW}\big]\Big] = (0, \ldots, 0).$$

We have therefore,

$$\mathbb{V}[NPV_t] = \mathbb{V}[X_t \cdot {}^{(2)}\beta_t + v_t] = \mathbb{V}[X_t \cdot {}^{(2)}\beta_t] + \mathbb{V}[v_t]$$

$$\mathbb{V}[NPV_t] = \mathbb{V}[X_t \cdot {}^{(2)}\beta_t] + {}^v\sigma_t^2.$$

### 4.1.3. Equivalence of the optimal parameters

Let us now consider the two linear models associated to regressions (1) and (3), and the additional linear model on the unobserved variable $NAV_t$ associated to regression (5), defined as follows:

$$NAV_t = X_t \cdot {}^{(3)}\beta_t + w_t \quad (5)$$

In this model, the variable $NAV_t$ is approximated by its conditional expectation given the $\sigma$-field generated by the regressors $X_t$, considered as a linear combination of the regressors.

The underlying assumption of this model can therefore be written

$$\exists\, ^{(3)}\beta_t, \mathbb{E}[NAV_t|X_t] = X_t \cdot {}^{(3)}\beta_t.$$

We shall make the following assumption (satisfied in practice)

$$\bar{H}: \mathbb{E}[X_t^{tr} \cdot X_t] \text{ exists and is invertible}$$

Considering the three linear models and assuming $\bar{H}$, we have the following result, that will be considered in Subsection 4.2,

$$^{(1)}\beta_t = {}^{(2)}\beta_t = {}^{(3)}\beta_t. \qquad (6)$$

Indeed,

$$\mathbb{E}[\widetilde{NAV}_t \,|X_t] = X_t \,.\, {}^{(1)}\beta_t$$

Implies that

$$\mathbb{E}[X_t. u_t] = (0, \ldots, 0) \Leftrightarrow \mathbb{E}[X_t^{tr}.(\widetilde{NAV}_t - X_t \,.\, {}^{(1)}\beta_t)] = \begin{pmatrix} 0 \\ \vdots \\ 0 \end{pmatrix}$$

$$\Leftrightarrow \mathbb{E}[X_t^{tr}.\widetilde{NAV}_t\,] = \mathbb{E}[X_t^{tr}.X_t]^{(1)}\beta_t.$$

Because $X_t$ is $\mathcal{F}_t^{RW} - measurable$

$$\mathbb{E}[\widetilde{NAV}_t \,|X_t] = X_t \,.\, {}^{(1)}\beta_t \Rightarrow {}^{(1)}\beta_t = \mathbb{E}[X_t^{tr}.X_t]^{-1}.\mathbb{E}[X_t^{tr}.\widetilde{NAV}_t\,] = \mathbb{E}[X_t^{tr}.X_t]^{-1}.\mathbb{E}\left[X_t^{tr}.\mathbb{E}[\widetilde{NAV}_t \,|\mathcal{F}_t^{RW}]\right]$$

$$^{(1)}\beta_t = \mathbb{E}[X_t^{tr}.X_t]^{-1}.\mathbb{E}[X_t^{tr}.NAV_t] = {}^{(3)}\beta_t.$$

We obtain result (6) by considering the same demonstration, adapted to (3).

## 4.2. Comparison of the Curve Fitting and Least Squares Monte-Carlo approaches

### 4.2.1. Comparison Curve Fitting vs. LSMC

As for (2) and (4), we can find a similar formula for the model associated to regression (5), considering an adapted notation,

$$^{NAV}\sigma_t^2 = \mathbb{V}[X_t. {}^{(3)}\beta_t] + {}^w\sigma_t^2. \qquad (7)$$

With ${}^w\sigma_t^2$, the residuals variance. Eventually, we can derive:

$$\begin{cases} ^{NAV}\sigma_t^2 + \dfrac{^{NPV}\sigma_t^2}{P} = \mathbb{V}[X_t. {}^{(3)}\beta_t] + {}^u\sigma_t^2 \\ ^{NAV}\sigma_t^2 + {}^{NPV}\sigma_t^2 = \mathbb{V}[X_t. {}^{(3)}\beta_t] + {}^v\sigma_t^2 \end{cases},$$

and

$$^v\sigma_t^2 = {}^u\sigma_t^2 + \frac{P-1}{P}{}^{NPV}\sigma_t^2. \qquad (8)$$

Under the standard OLS assumption, both Curve Fitting and $LSMC$ estimators converge towards ${}^{(3)}\beta_t$ with respective asymptotic speed of convergence ${}^u\sigma_t\sqrt{\dfrac{\mathbb{E}(X_t^{tr}.X_t)^{-1}}{N'}}$ and ${}^v\sigma_t\sqrt{\dfrac{\mathbb{E}(X_t^{tr}.X_t)^{-1}}{N'}}$. For a more detailed presentation of the underlying assumptions, see Saporta (2011).

Let ${}^{CF}N$ be the number of scenarios considered to obtain the Curve Fitting estimator and ${}^{LSMC}N$ be the number of scenarios considered to obtain the Least Squares Monte-Carlo estimator. The implementation complexity of both Curve Fitting and $LSMC$ can be approximated by the number of $NPV$ necessary to implement these methodologies, that is to say exactly ${}^{LSMC}N$ for $LSMC$ and ${}^{CF}N \times P$ for Curve Fitting.

We now evaluate the level of $^{LSMC}N$ allowing to obtain an equal asymptotic speed of convergence for each methodology, given a fixed value for $^{CF}N \times P$, that is to say

$$^{u}\sigma_t \sqrt{\frac{\mathbb{E}(X_t^{tr}.X_t)^{-1}}{^{CF}N}} = {}^{v}\sigma_t \sqrt{\frac{\mathbb{E}(X_t^{tr}.X_t)^{-1}}{^{LSMC}N}},$$

which is equivalent to

$$\frac{^{u}\sigma_t^2}{^{CF}N} = \frac{^{v}\sigma_t^2}{^{LSMC}N}.$$

From (7) and (8) we have

$$^{LSMC}N = {}^{CF}N \times \left(1 + \frac{P-1}{P}\frac{^{NPV}\sigma_t^2}{^{u}\sigma_t^2}\right) = {}^{CF}N \times \left(1 + \frac{P-1}{P}\frac{^{NPV}\sigma_t^2}{^{NAV}\sigma_t^2 + \frac{^{NPV}\sigma_t^2}{P} - \mathbb{V}[X_t.^{(3)}\beta_t]}\right),$$

and therefore,

$$^{LSMC}N = {}^{CF}N \times P \times \left(\frac{1 + \frac{^{w}\sigma_t^2}{^{NPV}\sigma_t^2}}{1 + P \times \frac{^{w}\sigma_t^2}{^{NPV}\sigma_t^2}}\right) \leq {}^{CF}N \times P. \quad (9)$$

In particular, we obtain the equality $^{LSMC}N = {}^{CF}N$ when $P = 1$, which is obvious by definition of $\widehat{NAV}_t$. According to (9) the Least Squares Monte-Carlo approach seems more efficient than the Curve Fitting methodology. The speed of convergence of its estimator is indeed better than or equal to the Curve Fitting estimator's speed of convergence, for an equivalent algorithmic complexity. This formula enables to compare the efficiency of the $LSMC$ and the Curve Fitting approach where the comparative efficiency coefficient

$$\eta = \sqrt{\frac{1 + \frac{^{w}\sigma_t^2}{^{NPV}\sigma_t^2}}{1 + P \times \frac{^{w}\sigma_t^2}{^{NPV}\sigma_t^2}}} \text{ appears.}$$

Considering result (7), we know that $^{w}\sigma_t^2 \in [0; {}^{NAV}\sigma_t^2]$. Eventually, both formulas obtained for the endpoints of this interval are actually intuitive.

### 4.2.2. Comparison of extreme cases

#### 4.2.2.1. Extreme case $^{w}\sigma_t^2 = 0$

This case happens when the hidden variable $NAV_t$ can be rewritten exactly as a polynomial function of the underlying risk factors which is the assumption that basically justifies the use of polynomial proxies.

If $^{w}\sigma_t^2 = 0$ then $^{LSMC}N = {}^{CF}N \times P$. This means that, if regression (5) is fully efficient, both procedures $LSMC$ and Curve Fitting have the same asymptotic speed of convergence for an equal algorithmic complexity.

As excellent results are generally obtained by using polynomial proxies, we can reasonably assume that $^{w}\sigma_t^2$ is close to 0, which leads to $^{w}\sigma_t^2 \ll {}^{NPV}\sigma_t^2$ and $\eta$ is close to 1. Then both approaches Curve Fitting and $LSMC$ asymptotically converge with a similar speed.

#### 4.2.2.2. Extreme case $^{w}\sigma_t^2 = {}^{NAV}\sigma_t^2$.

In this scenario the comparative efficiency coefficient takes its lowest possible value. This characterizes the total inefficiency of the replication of variable $NAV_t$ by the regression $X_t.^{(3)}\beta_t$. In this framework we have

$${}^{(3)}\beta_t = \begin{pmatrix} \mathbb{E}[NAV_t] \\ 0 \\ \vdots \\ 0 \end{pmatrix}.$$ Therefore, the difference between both asymptotic speeds of convergence is entirely explicated by the difference of the asymptotic speed of convergence of the estimators of $\mathbb{E}[NAV_t]$ obtained from the samples $\left(\widehat{NAV}_t^{n'}\right)_{n' \in [\![1; {}^{CF}N]\!]}$ and $\left(NPV_t^{n'}\right)_{n' \in [\![1; {}^{LSMC}N]\!]}$ used in the calibration steps.

Considering the Central Limit Theorem, the estimators $\frac{1}{{}^{CF}N}\sum_{n'=1}^{{}^{CF}N} \widehat{NAV}_t^{n'}$ and $\frac{1}{{}^{LSMC}N}\sum_{n'=1}^{{}^{LSMC}N} NPV_t^{n'}$ asymptotically converges towards Gaussian distributions that have a same mean $\mathbb{E}[NAV_t]$ and respective volatilities $\sqrt{\frac{{}^{NAV}\sigma_t^2 + \frac{{}^{NPV}\sigma_t^2}{P}}{{}^{CF}N}}$ and $\sqrt{\frac{{}^{NAV}\sigma_t^2 + {}^{NPV}\sigma_t^2}{{}^{LSMC}N}}$.

Equalizing both asymptotic speeds of convergence we obtain

$${}^{LSMC}N = {}^{CF}N \times P \left( \frac{1 + \frac{{}^{NAV}\sigma_t^2}{{}^{NPV}\sigma_t^2}}{1 + P \times \frac{{}^{NAV}\sigma_t^2}{{}^{NPV}\sigma_t^2}} \right).$$

Which is the same formula as (9), with ${}^w\sigma_t^2 = {}^{NAV}\sigma_t^2$. This observation enables us to validate formula (9) a posteriori, in this specific case.

### 4.2.3. Conclusion of this comparison section

Under our assumptions the $LSMC$ approach seems quite more efficient asymptotically than the Curve Fitting, for a fixed algorithmic complexity. However we must put this result in perspective. Indeed, for example, if we consider that $\mathcal{F}_t^{RW} = \sigma(X_t)$, which can be the case if $\{ {}^s\varepsilon_1, {}^{ZC}\varepsilon_1, \ldots, {}^s\varepsilon_t, {}^{ZC}\varepsilon_t \} \subset \sigma(X_t)$, our assumptions lead us to the result

$$NAV_t = \mathbb{E}[NAV_t | \mathcal{F}_t^{RW}] = \mathbb{E}[NAV_t | X_t] = X_t . {}^{(3)}\beta_t.$$

This situation corresponds to the first extreme case ${}^w\sigma_t^2 = 0$ for which both approaches $LSMC$ and Curve Fitting have equal efficiency for an equal algorithmic complexity.

In practice the case $\mathcal{F}_t^{RW} = \sigma(X_t)$ only happens for small values of $t$ ($X_t$ contains a limited number of covariates, when $t$ is too high we have $\sigma(X_t) \subset \mathcal{F}_t^{RW}$ strictly). Considering approximations of ${}^w\sigma_t^2$ and ${}^{NPV}\sigma_t^2$ calculated with traditional Asset-Liability Management models under current economic conditions, we have obtained orders of magnitude for $\eta$, varying increasingly from around 1 at $t = 1$ to 0.5 at $t = 5$. In this last case, the $LSMC$ asymptotical speed of convergence is 2 times greater than the Curve Fitting.

As a conclusion on (9), one operational complexity element is omitted in the formula. The algorithmic complexity is limited to the number of $NPV$ calculated to implement both approaches (${}^{LSMC}N$ for the $LSMC$ and ${}^{CF}N \times P$ for the Curve Fitting). However, another complexity factor may be relevant, the implementation time of the methodologies. Indeed, the $LSMC$ proxies are longer to calibrate it requires a large number of outcomes in its calibration step compared to Curve Fitting (in practice ${}^{LSMC}N$ is close to $P$ times ${}^{CF}N$). The impact of this complexity element is difficult to grasp because it highly depends on the calibration software and on the complexity of the underlying insurance products. Thus we have decided to get around this issue and consider a similar implementation time for both approaches.

# 5. Illustration

## 5.1. Implementation framework

The multi-year Curve Fitting and Least Squares Monte-Carlo methodologies are tested on a standard French saving portfolio with low average guaranteed minimum rates (0.81% on average but about 30% of the contracts are 2.5% or more guaranteed rates). We use a projection tool that takes profit sharing mechanisms, target crediting rate and dynamic lapses behaviors of policy holders into account. The credited rate paid to policy holders is a function of the risk-free rate, of the performance of the Eurostoxx index, and of a specific profit sharing rule. The dynamic lapses depend on the credited rate and on a market based target crediting rate. Basically, a low credited rate will lead to a high number of redeemed contracts. The asset reallocation rule is such that the initial asset allocation in maintained over the projections.

Table 1: Initial asset allocation (Market value)

| Asset | Allocation |
|---|---|
| Stock | 14% |
| Real Estate | 3% |
| Bonds | 78% |
| Cash | 5% |

The tested product is supposed to be subjected to a risk on the stock index and on interest rates.[9] This leads us to consider one stock risk factor and one interest rate risk factor. The time horizon selected is $T = 5$ years. The economic scenarios tables are calibrated using a specific calibration process is considered for the interest rates model, as developed in 5.2. The other economic assumptions have been calibrated as at the 31/12/2009, and not as at the 31/12/2010 or 31/12/2011 in order to avoid simulatory aberrations linked to very low interest rates and very high market implied volatilities. Basically, the calibration date has little impact on the implementation and this issue has only a secondary interest for our study.

The parameters of the polynomial proxies depend on the chosen initial assumptions. The change of one hypothesis (asset allocation, economic information, ALM rules…) implies anew calibration of the proxies.

## 5.2. Real-World interest rates modeling issue

The 1-year horizon considered when calculating the Economic Capital (Pillar I) allows us to consider a standard Real-World model like the 1-factor Heath-Jarrow-Morton. This model integrates historical risk premiums and volatilities per maturity in the Zero-Coupon bonds generation process. For a more developed insight about the Heath-Jarrow-Morton model, the reader may consult Brigo and Mercurio (2001).

Considering the same historical parameters throughout a 5-years horizon can lead to a significant issue when the initial interest rates are low. In particular, the introduction of countercyclicality in the interest rate model is legitimate in the context of low interest rates as at the 31/12/2009. Basically, this is not a major issue for our implementations. We have gotten around it by calibrating the risk premiums so as to obtain on average a specific long-term curve in $t = 5$, the Zero-Coupon bond curve as at the 31/12/2005.

We have been able to verify that this choice has little impact on the efficiency of the methodologies implemented hereafter. Other approaches based on time series theory or Principal Component Analysis, such as the methodology presented in Diebold and Li (2006), may lead to interesting solutions for this issue.

---

[9] We only focus in this article on the level risk and do not consider the volatility risk.

### 5.3. Multi-year Curve Fitting implementation

We have carried out Nested Simulations projections based on 5'000 real world simulations from $t = 1$ to $t = 5$. For each primary simulation and date $t$, the $NAV$ (central and shocked) are estimated with adjusted risk-neutral simulations based on a secondary table consisting of 500 scenarios. For more insight about the adjustment of risk-neutral simulations in the Nested Simulations framework, see Devineau (2010). This implementation has enabled us to obtain a reference set of joint sample paths $(\widehat{NAV}_1^n, \ldots, \widehat{NAV}_5^n)_{n \in [\![1;5'000]\!]}$ and $(\widehat{SCR}_1^n, \ldots, \widehat{SCR}_5^n)_{n \in [\![1;5'000]\!]}$. The study of the goodness of fit between the distributions obtained thanks to our proxies (Curve Fitting and Least Squares Monte-Carlo) and the Nested Simulations distributions enables to validate the calculations at each step of the process.

The multi-year Curve Fitting has been calibrated on 100 outcomes of Nested Simulations $NAV$ outcomes. This low number of outcomes enables us to implement the Nested Simulations procedure 50 times faster.

Two different approaches of the Curve Fitting methodology have been tested, implementing a calibration on well-dispersed primary scenarios, relying on a multidimensional risk factors' grid (obtained using multidimensional Sobol sequences), and a calibration on extreme primary scenarios. In this latter case, the calibration scenarios have been selected according to the high values of a norm on the risk factors,

$$\forall t \in [\![1;5]\!], \left\| \left( {}^s\varepsilon_1, {}^{ZC}\varepsilon_1, \ldots, {}^s\varepsilon_t, {}^{ZC}\varepsilon_t \right) \right\| = \sqrt{\sum_{u=1}^{t} \left( {}^s\varepsilon_u{}^2 + {}^{ZC}\varepsilon_u{}^2 \right)}.$$

Considering this norm, we can select scenarios whose risk factors vectors are associated to low probability thresholds. This type of scenarios' selection is already used to accelerate the calculation of the Economic Capital as in Devineau and Loisel (2009). Both approaches have led to satisfactory results. The results presented in Subsection 5.3 have been obtained by using the second approach.[10]

### 5.3.1. QQ plots

During former tests, we have been able to check that in most cases the final form of the optimal central and shocked proxies were very close if not the same. Therefore, in both the Curve Fitting and $LSMC$ implementation, we have chosen not to integrally duplicate the procedure described for central $NAV$ when calibrating the shocked proxies. Instead, we have considered the same set of regressors for the shocked proxies as those already calibrated for the central proxies. This has enabled us to accelerate the proxy methodologies implementation by skipping one part of the calibration process for the shocked proxies. The following graphs depict the quantile-quantile plots that compare the quantiles of the reference sets of $NAV_t$ with the quantiles of the sets obtained using the Curve Fitting methodology.

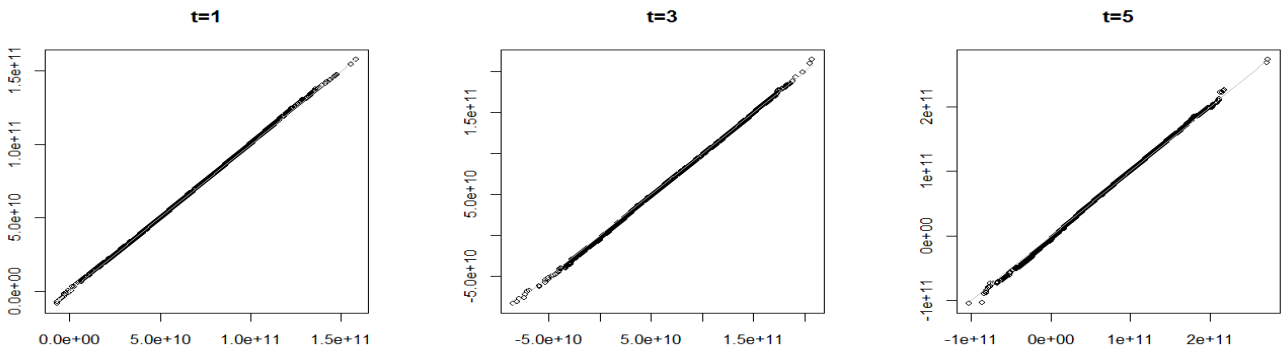

Figure 8: QQ plots $\widehat{NAV}_t$ vs. ${}^{CF}NAV_t$

---

[10] For more visibility, a little number of points of the top right corners of the QQ plots associated to the Solvency Ratio have been removed from the graphs. These points correspond to the more positive cases and are of little interest here.

Denoting $\widehat{SR}_t$ the solvency ratio obtained by Nested Simulations and $^{CF}SR_t$ the solvency ratio obtained by the multi-year Curve Fitting approach, we obtain the following QQ plots.

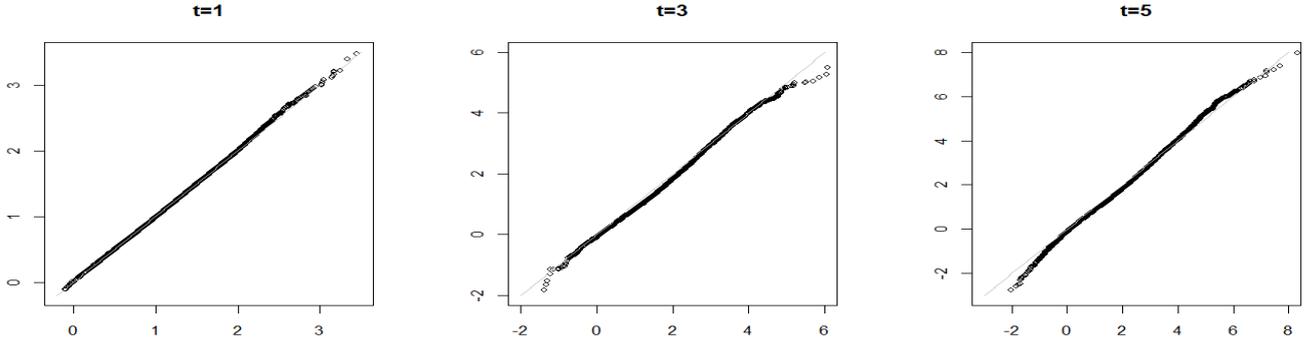

Figure 9: QQ plots $\widehat{SR}_t$ vs. $^{CF}SR_t$

The Solvency Ratio is a more complex mathematical element and one can observe that it is not as well approximated as the $NAV$. However the QQ plots obtained show satisfactory results for both approximated variables even for $t = 5$.

### 5.3.2. Relative differences between Nested Simulations and Curve Fitting

The relative differences observed between the Nested Simulations outcomes and the Curve Fitting approximations for chosen quantiles are shown below.

|          | NAV     |         |         | SR      |         |         |
|----------|---------|---------|---------|---------|---------|---------|
| Quantile | t=1     | t=3     | t=5     | t=1     | t=3     | t=5     |
| 25%      | 0.63%   | 4.11%   | 1.72%   | 1.62%   | 22.05%  | 15.25%  |
| 50%      | 0.08%   | 2.65%   | -0.59%  | 1.08%   | 13.77%  | 8.71%   |
| 75%      | -0.01%  | 2.09%   | -0.45%  | 0.33%   | 6.18%   | 1.20%   |

Very low relative differences can be observed as far as the $NAV$ approximation is concerned. The relative differences in respect of the Solvency Ratio are much higher, especially for the lowest quantiles, when the outcomes are closer to 0. This was somehow predictable because the Solvency Ratio is a much more complex element than the $NAV$. As one can see further, the $LSMC$ implementation led to better relative differences.

### 5.3.3. Final shape of the calibrated proxies

The calibrated "central" proxies present the following shapes.

$$^{CF}NAV_1 = I + \alpha_1\ ^s\varepsilon_1 + \alpha_2\ ^{ZC}\varepsilon_1 + \alpha_3\ ^s\varepsilon_1{}^2 + \alpha_4\ ^{ZC}\varepsilon_1{}^2 + \alpha_5\ ^s\varepsilon_1 \cdot {}^{ZC}\varepsilon_1 + \alpha_6\ ^s\varepsilon_1{}^3 + \alpha_7\ ^s\varepsilon_1{}^2 \cdot {}^{ZC}\varepsilon_1 + \alpha_8\ ^s\varepsilon_1 \cdot {}^{ZC}\varepsilon_1{}^2,$$

$$^{CF}NAV_3 = I + \alpha_1\ ^{CF}NAV_2 + \alpha_2\ ^s\varepsilon_3 + \alpha_3\ ^{ZC}\varepsilon_3 + \alpha_4\ ^{ZC}\varepsilon_3{}^2 + \alpha_5\ ^s\varepsilon_3 \cdot {}^{ZC}\varepsilon_3 + \alpha_6\ ^s\varepsilon_3 \cdot {}^s\varepsilon_2 + \alpha_7\ ^s\varepsilon_3 \cdot {}^s\varepsilon_1 \\ + \alpha_8\ ^{ZC}\varepsilon_3 \cdot {}^{ZC}\varepsilon_2 + \alpha_9\ ^{ZC}\varepsilon_3 \cdot {}^{ZC}\varepsilon_1 + \alpha_{10}\ ^{ZC}\varepsilon_2 + \alpha_{11}\ ^{ZC}\varepsilon_1,$$

$$^{CF}NAV_5 = I + \alpha_1\ ^{CF}NAV_4 + \alpha_2\ ^s\varepsilon_5 + \alpha_3\ ^{ZC}\varepsilon_3 + \alpha_4\ ^s\varepsilon_5{}^2 + \alpha_5\ ^{ZC}\varepsilon_5{}^2 + \alpha_6\ ^s\varepsilon_5 \cdot {}^s\varepsilon_3 + \alpha_7\ ^{ZC}\varepsilon_5 \cdot {}^{ZC}\varepsilon_4 + \alpha_8\ ^{ZC}\varepsilon_5 \cdot {}^{ZC}\varepsilon_3 \\ + \alpha_9\ ^{ZC}\varepsilon_5 \cdot {}^{ZC}\varepsilon_2 + \alpha_{10}\ ^{ZC}\varepsilon_5 \cdot {}^{ZC}\varepsilon_1 + \alpha_{11}\ ^{ZC}\varepsilon_1.$$

One can interpret the proxies shapes the following way. The first order regressors reflects the linear sensibility of the $NAV$ to the risk. The second order terms permit to introduce the convexity of the $NAV$ and the higher order terms enables convexity adjustments. It is noticeable that the term $^{CF}NAV_{t-1}$ is always greatly significant in the regression associated to $^{CF}NAV_t$ ($t \geq 2$).

## 5.4. Multi-year LSMC implementation

The multi-year $LSMC$ has been calibrated on 50'000 outcomes of Net Present Value of margins. As for the multi-year Curve Fitting, two different approaches of the $LSMC$ methodology have been tested, considering a calibration on well-dispersed primary scenarios or a calibration on extreme primary scenarios. Both methods have led to very satisfactory results. The results presented in this section have been obtained using the approach based on extreme scenarios.[11]

### 5.4.1. QQ plots

The following graphs depict the QQ plots that compare the quantiles of the reference sets of $NAV_t$ with the quantiles of the sets obtained using the LSMC methodology.

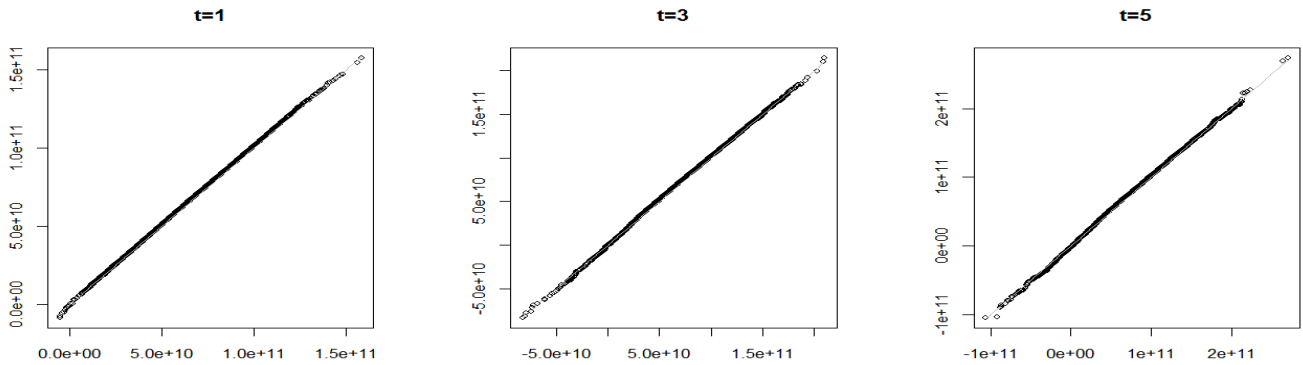

Figure 10: QQ plots $\widehat{NAV}_t$ vs. $^{LSMC}NAV_t$

Denoting $^{LSMC}SR_t$ the solvency ratio obtained by the multi-year $LSMC$ approach, we obtain the following QQ plots.

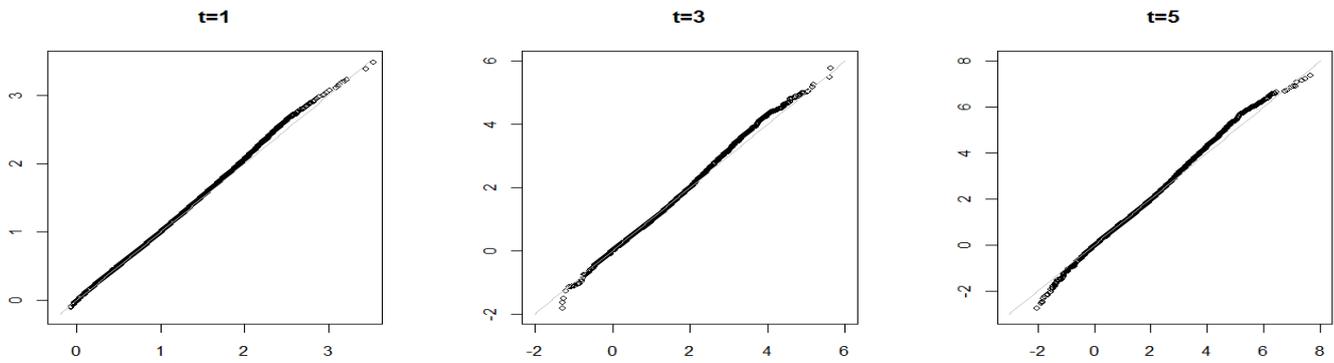

Figure 11: QQ plots $\widehat{SR}_t$ vs. $^{LSMC}SR_t$

Similarly to the Curve Fitting results, the Solvency Ratio is not as well approximated as the $NAV$. However the QQ plots obtained show satisfactory results for both approximated variables event for $t = 5$.

---

[11] For more visibility, few points of the top right corner of the Q – Q plots associated to the Solvency Ratio have been removed from the graphs. These points correspond to the best cases and are of little interest here.

### 5.4.2. Relative differences between Nested Simulations and LSMC

We summarize below the relative differences observed between the Nested Simulations outcomes and the Curve Fitting approximations for chosen quantiles.

| | NAV | | | SR | | |
|---|---|---|---|---|---|---|
| **Quantile** | t=1 | t=3 | t=5 | t=1 | t=3 | t=5 |
| **25%** | -1.92% | -3.58% | -5.51% | 0.76% | 8.66% | 4.80% |
| **50%** | -1.79% | -2.63% | -2.47% | 0.03% | 3.15% | 2.60% |
| **75%** | -1.39% | -1.28% | -0.79% | -1.42% | -1.95% | -4.75% |

Very low relative differences can be observed for both the $NAV$ and the Solvency Ratio, even for the lowest quantiles.

### 5.4.3. Final shape of the calibrated polynomial proxies

The calibrated proxies present the following shapes.

$$^{LSMC}NAV_1 = I + \alpha_1 \, {}^s\varepsilon_1 + \alpha_2 \, {}^{ZC}\varepsilon_1 + \alpha_3 \, {}^s\varepsilon_1{}^2 + \alpha_4 \, {}^{ZC}\varepsilon_1{}^2 + \alpha_5 \, {}^s\varepsilon_1 . {}^{ZC}\varepsilon_1 + \alpha_6 \, {}^s\varepsilon_1{}^3 + \alpha_7 \, {}^s\varepsilon_1{}^2 . {}^{ZC}\varepsilon_1 ,$$

$$^{LSMC}NAV_3 = I + \alpha_1 \, {}^{LSMC}NAV_2 + \alpha_2 \, {}^{LSMC}NAV_2 . {}^{ZC}\varepsilon_3 + \alpha_3 \, {}^s\varepsilon_3 + \alpha_4 \, {}^{ZC}\varepsilon_3 + \alpha_5 \, {}^s\varepsilon_3{}^2 + \alpha_6 \, {}^{ZC}\varepsilon_3{}^2 + \alpha_7 \, {}^s\varepsilon_3 . {}^{ZC}\varepsilon_3 \\ + \alpha_8 \, {}^s\varepsilon_3 . {}^s\varepsilon_2 + \alpha_9 \, {}^s\varepsilon_3 . {}^s\varepsilon_1 + \alpha_{10} \, {}^{ZC}\varepsilon_3 . {}^{ZC}\varepsilon_2 + \alpha_{11} \, {}^{ZC}\varepsilon_3 . {}^{ZC}\varepsilon_1 + \alpha_{12} \, {}^{ZC}\varepsilon_2 ,$$

$$^{LSMC}NAV_5 = I + \alpha_1 \, {}^{LSMC}NAV_4 + \alpha_2 \, {}^{LSMC}NAV_4 . {}^s\varepsilon_5 + \alpha_3 \, {}^s\varepsilon_5 + \alpha_4 \, {}^{ZC}\varepsilon_3 + \alpha_5 \, {}^s\varepsilon_5{}^2 + \alpha_6 \, {}^{ZC}\varepsilon_5{}^2 + \alpha_7 \, {}^{ZC}\varepsilon_5 . {}^{ZC}\varepsilon_4 \\ + \alpha_8 \, {}^{ZC}\varepsilon_5 . {}^{ZC}\varepsilon_3 + \alpha_9 \, {}^{ZC}\varepsilon_5 . {}^{ZC}\varepsilon_2 + \alpha_{10} \, {}^{ZC}\varepsilon_5 . {}^{ZC}\varepsilon_1 + \alpha_{11} \, {}^{ZC}\varepsilon_1 .$$

One can observe that both Curve Fitting and $LSMC$ proxies have very similar shapes. Basically, the interpretation of the proxies' shapes is similar to the one developed in 5.3.3. Once again the term $^{CF}NAV_{t-1}$ is always highly significant in the regression associated to $^{CF}NAV_t$ ($t \geq 2$).

## 5.5. Calculation of the capital need associated to a constraint on solvency shortfalls

The solvency constraints have been chosen for their realism, knowing that the initial Solvency Ratio for the considered product, calculated using the Standard Formula, is close to 90%. We have considered two different solvency constraints, one on yearly Solvency Ratio distributions and one on Solvency Ratio's paths.

| **Multi-year solvency constraint** | **LSMC** | **Curve Fitting** |
|---|---|---|
| **(SC3)** : $\forall t \in [|1;T|], \mathbb{P}\left(\frac{NAV_t}{SCR_t} \geq 110\%\right) \geq 85\%$. | 7.9% | 6.7% |
| **(SC4)** : $\mathbb{P}\left(\cap_{t=1}^{T}\left\{\frac{NAV_t}{SCR_t} \geq 110\%\right\}\right) \geq 85\%$. | 9.3% | 11.9% |

One can observe that the relative difference between the capital need obtained by using both proxy methodologies and the one obtained by considering Nested Simulations projections, never exceeds 12%. This final result tends to legitimate the use of these methodologies in order to quickly approximate the Overall Solvency Needs related to these highly complex constraints. One can notice that the second constraint is framed such as it leads to a lower risk tolerance than the second constraint, which does not take path-dependence into account.

# Conclusion

The Own Risk and Solvency Assessment process implementation leads to numeral strategic issues for insurance undertakings. In this paper we have introduced the issues raised by the Overall Solvency Needs assessment and by the multi-year solvency concept. We have formalized various multi-year solvency metrics that can be used to provide a framework for the Overall Solvency Needs calculation Several modeling issues have been identified for the implementation of the most complex constraints. In order to work around these points we have developed multi-year proxy methodologies which have provided very interesting results, especially for the Least Squares Monte-Carlo methodology.

The authors notice that the modeling choices considered in this paper do not claim to model the risks in the most appropriate way, especially for long term risk management. However, they are generally considered by practitioners and therefore are relevant in our operational framework.

Concerning the future axes to investigate, it is relevant to address the issue of the proxies recalibration frequency needed to monitor the Overall Solvency Needs in an infra-annual fashion. Indeed, the proxies calibrations greatly depend on the economic assumptions and on the asset-mix at the implementation date.

Eventually, the framework considered in the illustration does not enable us to challenge empirically the efficiency comparison formula obtained in Section 4. This direction of work will be investigated further in order to analyze the robustness of our assumptions and to compare efficiently the respective speed of convergence of the proxy methodologies.

# Acknowledgement

The authors would like to thank Fabien Ramaharobandro who has actively contributed to this article and Stephane Loisel for his relevant comments throughout the writing of the paper.

The authors would like to extend their thanks to all the employees of Milliman Paris, and in particular the members of the R&D team.

Eventually, the authors would also like to express their gratitude to Milliman for having funded this project.